\documentclass[11pt]{article} 
\usepackage{graphicx,epsfig,caption,natbib}
\begin{document}

\title{The Gender Breakdown of the Applicant Pool for Tenure-Track Faculty Positions at a Sample of North American Research Astronomy Programs}
\author{Todd A. Thompson\\
Department of Astronomy\\
The Ohio State University\\
thompson@astronomy.ohio-state.edu}

\maketitle

\vspace*{-0.8cm}
\section{Introduction}
\label{section:introduction}
\vspace*{-0.3cm}

The demographics of the field of Astronomy is an active area of investigation.  Among many characteristics of the population, gender --- including gender balance, gender bias, and the gender-related component of the leaky pipeline --- have been the focus of recent work.  

The American Astronomical Society's Committee on the Status of Women (CSWA)\footnote{http://www.aas.org/cswa/} published the results of a survey in 2013\footnote{http://www.aas.org/cswa/status/Status\_2014\_Jan.pdf} that provides information on the fraction of women at each level in the astronomical workforce. For graduate students (at all levels) they report $F/(F+M)=404/1155 = 0.350$, while for postdoctoral researchers $F/(F+M)=186/645 = 0.288$, and for assistant professors $F/(F+M)=57/193= 0.295$.  

A piece of information missing from this report is the fraction of women in the faculty application pool.  I participated in several hiring committees for tenure-track faculty in Astronomy at Ohio State University in recent years, and I found that $F/(F+M)\simeq0.2$ among our applicants, significantly lower than the fraction of female postdoctoral researchers reported by the CSWA.  

This led to the question of whether or not OSU's number was typical of application pools for tenure-track positions at other research institutions.  To answer this question, and as a guide for other Departments and hiring committees, I contacted a number of institutions in an attempt to gather data on the gender breakdown of the applicant pool for tenure-track positions at research universities over the last few years. The result is presented in Figure \ref{figure:plot}.

\vspace*{-0.3cm}
\section{Methodology, Sample, \& Results}
\label{section:methodology}
\vspace*{-0.3cm}

I selected job searches from the AstroBetter Rumor Mill over the last few years.\footnote{http://www.astrobetter.com/wiki/tiki-index.php?page=Rumor+Mill+Faculty-Staff} All of the searches selected were advertised as tenure-track positions at research institutions (non-liberal arts).  All of the institutions selected are in the US except one. I contacted a total of approximately 30 institutions.  All were asked for the total number of applicants and the gender breakdown of the pool in their recent searches, and I told each that the information provided would be presented in anonymous/aggregate form. In a number of cases, institutions sent data on more than one search.  No cut was made for targeted searches (e.g., theory, instrumentation, observation, computation), or for Astronomy-only or combined Astronomy and Physics Departments, or on applications originating in the US or internationally, or whether or not the search was successful.  All searches reported here occurred after the 2010 job cycle and most are from 2012/2013 and 2013/2014. 

The way in which the numbers ($F$ and $F+M$) were calculated at each institution varied.  For example, in several cases the total number of applicants and the breakdown between females and males was reported by the HR department of the institution and included only applications received by the stated application deadline, and only for those individuals who reported their gender.  In other cases, individuals on the search committees or administrative personnel went through the application pool names by hand, and identified gender by name.  In some cases, names are ambiguous.  In all searches possible, ambiguous names were left out of the final tally. For these reasons, one should view the numbers presented below with an associated uncertainty.  In cases where we could make an estimate of this uncertainty, it was $\simeq5$\% in the quantity $F/(F+M)$.  Clearly the methodology could be improved by normalizing the way in which $F/(F+M)$ was calculated for each search.

The final dataset includes 35 searches at 25 institutions.  A subset of the total sample is ranked in the 33 Astronomy Departments from the most recent National Research Council (NRC) report on graduate programs.\footnote{http://chronicle.com/article/NRC-Rankings-Overview-/124705/ \label{foot:nrc}}  

The primary result of this study is the left panel of Figure \ref{figure:plot}, which shows a histogram of $F/(F+M)$ for the main sample (black) and the NRC Astronomy sample (red). The mean and median are 0.18 and 0.19, and 0.19 and 0.20 for the total sample and NRC Astronomy sample, respectively. Both distributions are narrow with an associated dispersion of $\simeq0.03-0.04$.  One might worry that the distribution in the left panel is artificially narrow because in a given year the same people are applying for the same jobs. However, the data span more than two job cycles, and the total number of applications varies significantly in the sample, as shown in the right panel.

\vspace*{-0.3cm}
\section{Discussion}
\label{section:discussion}
\vspace*{-0.3cm}

The primary result of this study is that for a sample of research institutions with tenure-track faculty positions open over the last few years the median value of the ratio $F/(F+M)\simeq0.2$, and that this ratio is fairly independent of the total number of applicants (Figure \ref{figure:plot}).

It is tempting to interpret Figure \ref{figure:plot} in the context of the values of $F/(F+M)$ reported by the CSWA for graduate students, postdocs, and assistant professors.  All are higher than any value of $F/(F+M)$ in the applicant pool for any search presented in Figure \ref{figure:plot}.  

Many interpretations of this fact are possible.  It is difficult to draw strong conclusions because of the incomplete nature of this study, and its methodological limitations.  For example, one might argue that because the fraction of assistant professors ($\simeq0.30$; \S1) is higher than in the applicant pool ($\simeq0.20$) that women are more likely than men to get a faculty position in Astronomy.  However, note that no cut has been made here for quality of application,\footnote{For example, taking the Hubble Postdoctoral Fellows for 2012, 2013, and 2014, I find that $F/(F+M)\simeq0.34$. See http://www.stsci.edu/institute/smo/fellowships/hubble/fellows-list/.}  or appropriateness of application given the institution and its stated focus/interests in its ad. One possibility is that men are more likely to send more applications to more places than female applicants, even if their research focus and qualifications are not commensurate with the advertised position.\footnote{See, for example, the study of gender differences in applying for promotions at Hewlett-Packard described in  http://www.theatlantic.com/features/archive/2014/04/the-confidence-gap/359815/.} Additionally, the faculty applicant pool may draw from the broader Physics community, where the demographics at all levels are different than Astronomy.  Finally, the fraction of female assistant professors is computed from US institutions, whereas the application pool draws from researchers worldwide.  The ratio of international/US applicants may be different than the ratio of both international/US postdoctoral researchers and assistant professors in the US.

Similar caveats apply in interpreting the higher $F/(F+M)$ ratio for postdoctoral researchers ($\simeq0.29$; \S1) relative to Figure \ref{figure:plot}.

Many of these issues could be studied by gathering more demographic information for each search. Comparison with the female/male fraction on the long shortlist, the short shortlist, and the fraction that eventually got the job would also be interesting, but has not been done here. A more extensive study could investigate the average time-dependence of $F/(F+M)$, the fluctuations from place to place in a given year, and the potential differences between programs with stand-alone Astronomy Departments and those with combined Physics and Astronomy Departments.  The gender breakdown for tenure-track faculty positions at liberal arts institutions would also be of interest.  

From the perspective of faculty search committees, Figure \ref{figure:plot} is useful regardless of the interpretation.  Other hiring committees will likely be interested to compare their own number in past and future searches to this distribution to decide whether or not they could be doing more as a Department to attract an applicant pool that is representative of the broader community.

\vspace*{-0.3cm}
\section{Acknowledgments}
\begin{figure}
\centerline{\includegraphics[width=8.5cm]{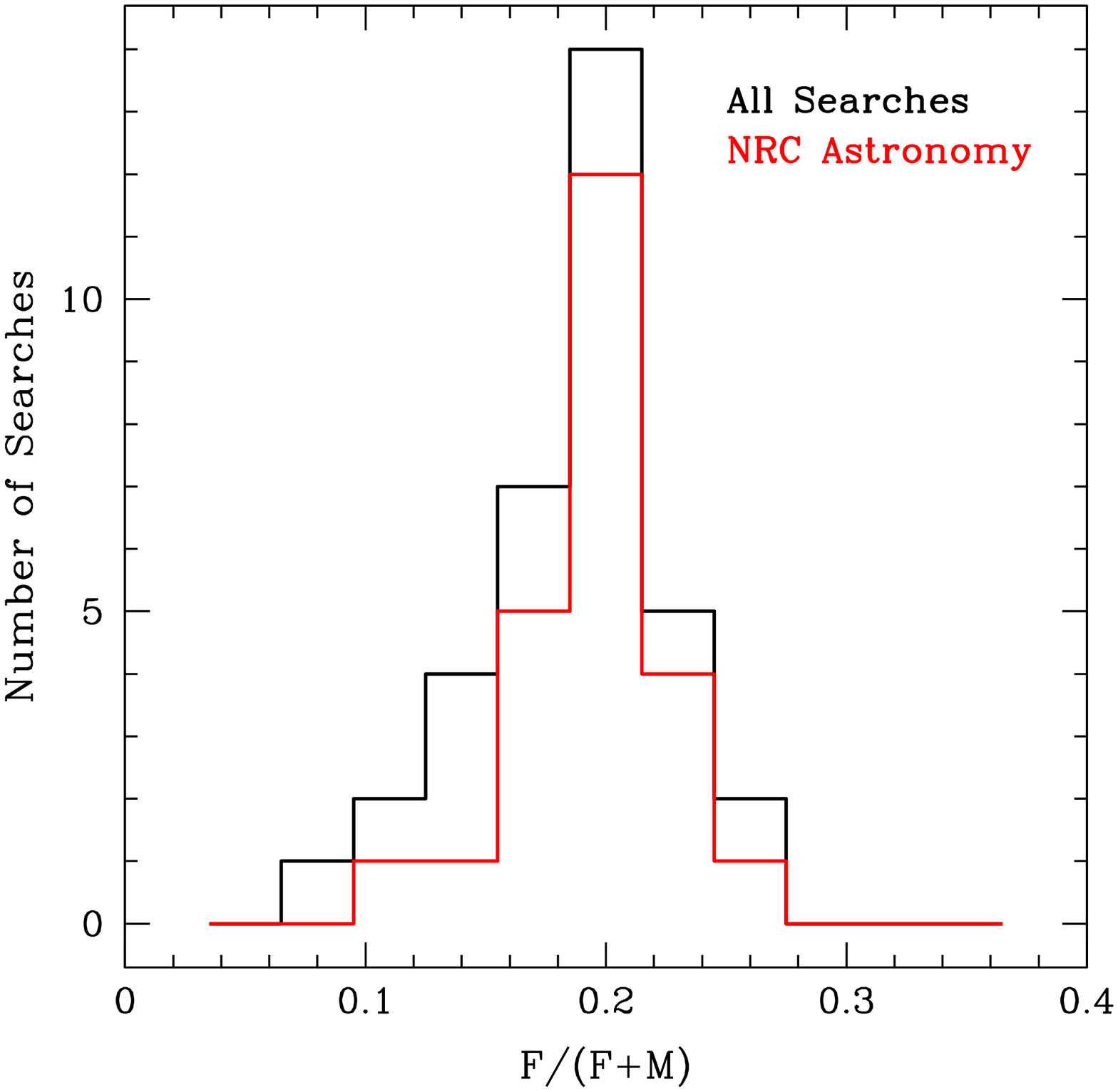}\includegraphics[width=8.5cm]{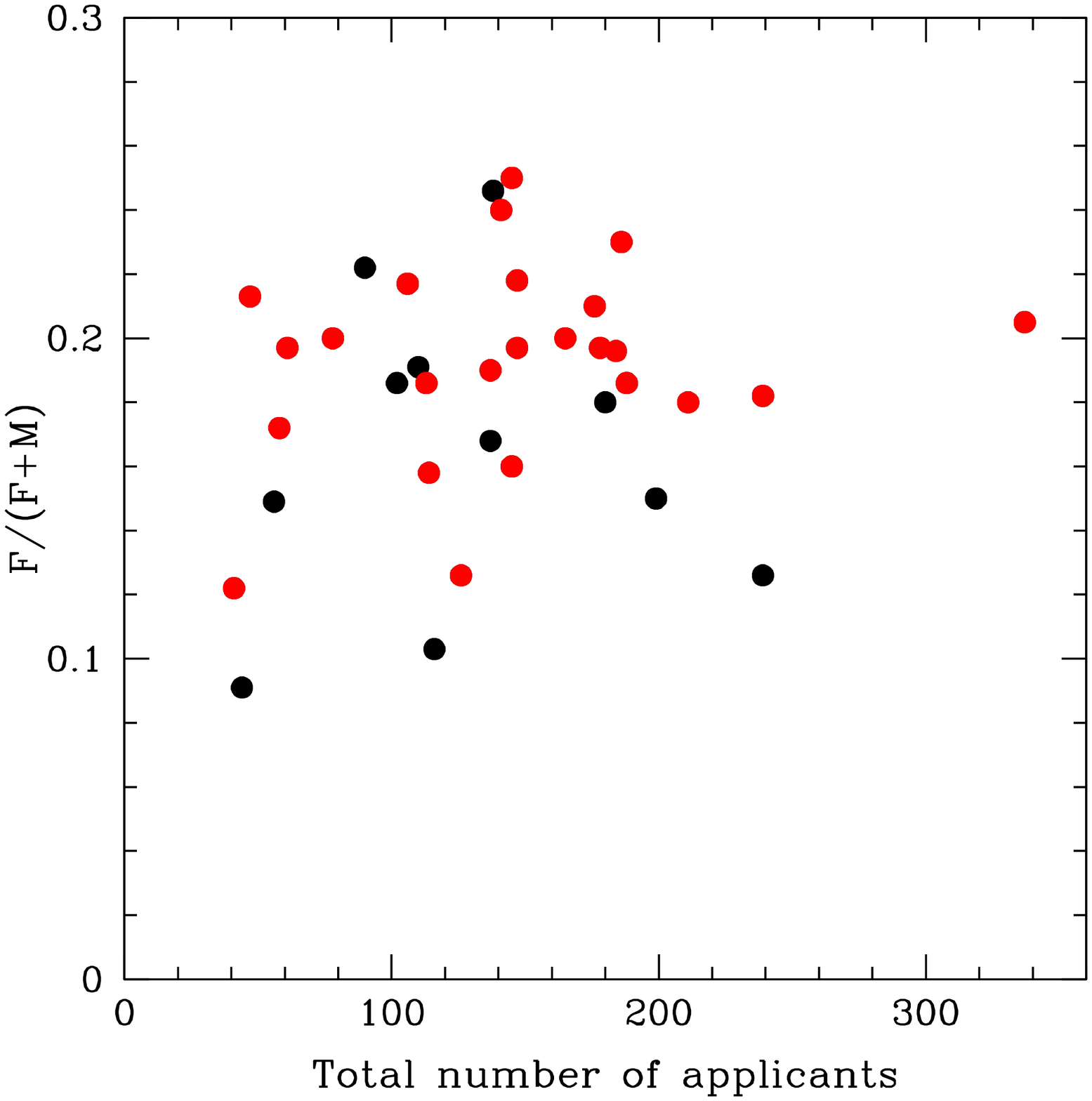}}
\caption{{\it Left:} Histogram of the number of searches versus $F/(F+M)$ in the total sample (black) and for the searches at NRC ranked Astronomy programs (red; see footnote \ref{foot:nrc}).  For the former, the mean and median of the distribution are $\simeq0.18\pm0.04$ and $0.19$, whereas for the latter they are $\simeq0.19\pm0.03$ and $0.20$. {\it Right:} $F/(F+M)$ versus the total number of applicants.
\label{figure:plot}}  
\end{figure}

\vspace*{-0.3cm}
I thank the many faculty and administrative staff who worked to provide the numbers shown in Fig.~\ref{figure:plot}.  I thank Annika Peter, Paul Martini, Katra Byram, Jennifer Johnson, Scott Gaudi, Brad Peterson, Meredith Hughes, Daryl Haggard, Joan Schmelz, John Beacom, and Lynne Hillenbrand  for useful conversations and suggestions.

\end{document}